%% file: main.tex
\documentclass[sigconf, pbalance, nonacm]{acmart}

\AtBeginDocument{%
  }

\usepackage{enumitem}
\usepackage{algorithm}
\usepackage{algorithmic}

\begin{document}

\title[SMART Retrieval]{SMART: LLM-Augmented Hybrid Retrieval for Dynamic Product Ads}

\settopmatter{authorsperrow=4, printacmref=true}

\author{Congfei Zhang}
\affiliation{\institution{Snap Inc.}\city{Bellevue}\state{WA}\country{USA}}
\email{czhang3@snapchat.com}

\author{Jingxiao Ma}
\affiliation{\institution{Snap Inc.}\city{Bellevue}\state{WA}\country{USA}}
\email{jma3@snapchat.com}

\author{Xiaodong Liu}
\affiliation{\institution{Snap Inc.}\city{Bellevue}\state{WA}\country{USA}}
\email{xliu9@snapchat.com}

\author{Hsiang-wei Chao}
\affiliation{\institution{Snap Inc.}\city{Seattle}\state{WA}\country{USA}}
\email{hchao@snapchat.com}

\author{Siman Wang}
\affiliation{\institution{Snap Inc.}\city{Bellevue}\state{WA}\country{USA}}
\email{swang7@snapchat.com}

\author{Ge Liu}
\affiliation{\institution{Snap Inc.}\city{Bellevue}\state{WA}\country{USA}}
\email{gliu@snapchat.com}

\author{Shantanu Aggarwal}
\affiliation{\institution{Snap Inc.}\city{Palo Alto}\state{CA}\country{USA}}
\email{saggarwal@snapchat.com}

\author{Vincent Zhang}
\affiliation{\institution{Snap Inc.}\city{Bellevue}\state{WA}\country{USA}}
\email{wzhang7@snapchat.com}

\author{Meghana Missula}
\affiliation{\institution{Snap Inc.}\city{Palo Alto}\state{CA}\country{USA}}
\email{mmissula@snapchat.com}

\author{Rachel Liao}
\affiliation{\institution{Snap Inc.}\city{New York}\state{NY}\country{USA}}
\email{rliao2@snapchat.com}

\author{Zichu Li}
\affiliation{\institution{Snap Inc.}\city{Palo Alto}\state{CA}\country{USA}}
\email{zli12@snapchat.com}

\author{Xiao Bai}
\affiliation{\institution{Snap Inc.}\city{Palo Alto}\state{CA}\country{USA}}
\email{xbai@snapchat.com}

\author{Yunzhi Zhou}
\affiliation{\institution{Snap Inc.}\city{Palo Alto}\state{CA}\country{USA}}
\email{yzhou10@snapchat.com}

\author{Yajun Wang}
\affiliation{\institution{Snap Inc.}\city{Palo Alto}\state{CA}\country{USA}}
\email{ywang30@snapchat.com}

\author{Zhe Liu}
\affiliation{\institution{Snap Inc.}\city{Palo Alto}\state{CA}\country{USA}}
\email{zliu11@snapchat.com}

\author{Jinchao Li}
\affiliation{\institution{Snap Inc.}\city{Bellevue}\state{WA}\country{USA}}
\email{jli18@snapchat.com}

\author{Yu Zhang}
\affiliation{\institution{Snap Inc.}\city{Palo Alto}\state{CA}\country{USA}}
\email{yzhang3@snapchat.com}

\renewcommand{\shortauthors}{Zhang et al.}

\begin{abstract}
Dynamic Product Ads (DPA) require retrieving relevant items from multi-million product catalogs, balancing two competing objectives: retargeting (re-surfacing known interests) and prospecting (discovering new categories). While Large Language Models (LLMs) capture semantic intent better than traditional embedding models, deploying them at scale introduces prohibitive inference costs and lexical mismatch issues. Through controlled experiments on millions of users, we demonstrate a critical retrieval decomposition: rule-generated queries excel at retargeting on a lexical BM25 index, while LLM-generated queries excel at prospecting on a dense ANN index. Building on this, we propose SMART (SeMantic-aware Adaptive ReTrieval). To manage costs, a lightweight quality gate identifies coverage gaps in initial keyword results, adaptively routing only the $\sim$10\% of users who benefit from semantic prospecting to the LLM path. Offline evaluation demonstrates that this gated approach captures the bulk of semantic prospecting gains in Relevance Score while maintaining competitive re-targeting performance at a 90\% reduction in LLM costs. Finally, in a 2-week online A/B test at Snap, SMART improved the ad conversion rate by $+$27.6\% over a strong embedding-based baseline.
\end{abstract}



\maketitle

\begin{figure*}[t]
    \centering
    \includegraphics[width=0.85\textwidth]{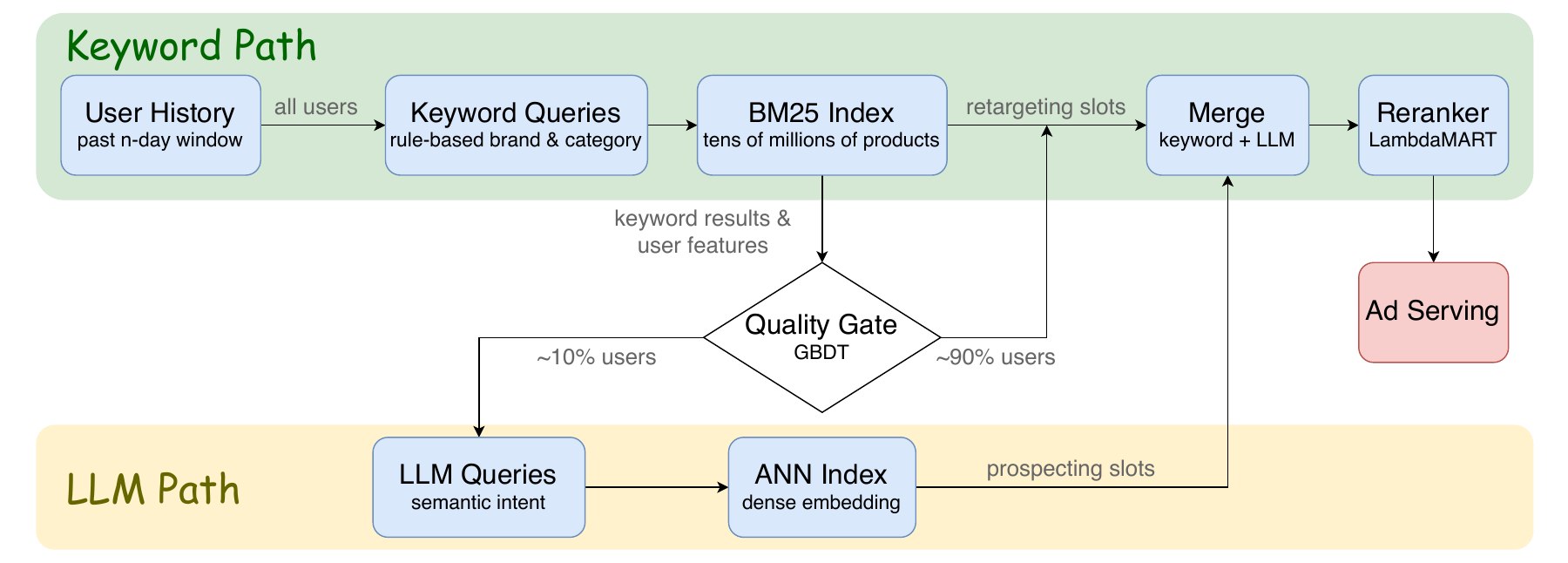}
    \caption{SMART system architecture. All users receive rule-generated queries for BM25 retargeting. The quality gate evaluates coverage gaps and routes $\sim$10\% of users to LLM prospecting queries on a dense ANN index. Candidates are merged, reranked, and sent to ad serving.}
    \label{fig:architecture}
\end{figure*}

\input{sections/introduction}

\input{sections/related_work}

\input{sections/preliminary_study}

\input{sections/architecture}

\input{sections/evaluation}

\input{sections/closing}

\bibliographystyle{ACM-Reference-Format}
\bibliography{references}

\end{document}

%% file: sections/introduction.tex
\section{Introduction}
\label{sec:intro}

Dynamic Product Ads (DPA) drive personalized recommendations by analyzing a user's shopping behavior. The core retrieval challenge involves sifting through millions of items to identify the most relevant products based on a user's browsing, cart, and purchase history. This task is uniquely difficult because it requires balancing two competing objectives. On one hand, systems must excel at retargeting---re-surfacing products from known brands and categories to drive immediate conversions. On the other hand, they must support prospecting---discovering relevant items in new categories to expand the user's consideration set.

Existing DPA retrieval systems rely heavily on collaborative filtering or embedding-based models~\citep{abutbul2023dpa, covington2016youtube}. While these approaches capture behavioral similarity, they struggle to express semantic intent (e.g., inferring ``running accessories'' from an athletic-footwear purchase). Large Language Models (LLMs) offer a promising alternative, as their pre-trained world knowledge allows them to interpret user histories and generate semantically rich search queries~\citep{zhu2023llm_retrieval, wang2023query2doc}. However, deploying LLMs in ad retrieval introduces three major challenges. First, there is a query--backend mismatch: LLM-generated text often fails to align with the strict lexical patterns required by traditional BM25 indexes~\citep{gao2022hyde, jagerman2023query}. Second, invoking an LLM for every user is cost-prohibitive and impractical under strict millisecond latency constraints~\citep{hu2026adnanny}. Finally, the industry lacks a controlled, production-scale comparison detailing when LLM-generated queries outperform simpler rule-based approaches.

To address these challenges, we propose SMART (SeMantic-aware Adaptive ReTrieval), which serves every user via a zero-cost rule-generated keyword pass and uses a lightweight quality gate to audit candidate coverage---invoking an LLM-driven dense retrieval path only for users left under-covered, as Figure~\ref{fig:architecture} illustrates. We show that rule-generated and LLM-generated queries are not substitutes, but serve distinct, complementary purposes. Rule-generated queries emit exact tokens from the user's history, making them ideal for retargeting on a lexical BM25 index. Conversely, LLMs are uniquely suited for semantic prospecting on a dense Approximate Nearest Neighbor (ANN) index, where their world knowledge surfaces complementary products that cannot be extracted from the user's history. This gated pairing achieves the best retrieval quality while reducing LLM costs by 90\%.

We dissect the efficacy of SMART through extensive offline experiments and a 2-week online A/B test against Snap's existing embedding-based baseline---a two-tower user-to-product (U2I) model with ANN retrieval. Our main contributions are:
\begin{itemize}[leftmargin=*]
    \item \textbf{A retargeting--prospecting decomposition:} We validate on a large-scale production cohort that rule-generated queries excel at retargeting (measured by CatRecall@$K$), while LLMs excel at prospecting (measured by Relevance Score).
    \item \textbf{A cross-metric evaluation:} We empirically demonstrate that CatRecall@$K$ and Relevance Score yield opposite rankings on the same systems, confirming that retargeting and prospecting necessitate different query strategies and backends.
    \item \textbf{A learned output-auditing quality gate:} We introduce a lightweight GBDT router that audits the output of the zero-cost keyword pass by measuring coverage gaps relative to user historical interests. By adaptively routing only the $\sim$10\% of users with the highest predicted coverage deficits to the LLM path, SMART captures peak retrieval quality while achieving a 90\% reduction in LLM inference costs.
    \item \textbf{Production impact:} We detail the deployed SMART system, which successfully routes rule-generated queries to BM25 and LLM queries to ANN, driving a $+$27.6\% relative improvement in conversion rate during a 2-week online A/B test at Snap.
\end{itemize}

%% file: sections/related_work.tex
\section{Related Work}
\label{sec:related}

\paragraph{LLM-Based Query Expansion.} Early generative query expansion techniques like Query2Doc~\citep{wang2023query2doc} append LLM-generated pseudo-documents to user queries, demonstrating consistent improvements on dense retrieval benchmarks but showing less reliable gains on sparse indices. Similarly, architectures like HyDE~\citep{gao2022hyde} generate hypothetical documents specifically for dense embedding vector spaces while avoiding lexical matching entirely due to term-frequency discrepancies. Comparative evaluations across sparse and dense retrievers confirm that LLM query expansions provide significantly larger and more stable performance lifts when paired with semantic backends~\citep{jagerman2023query}. Rather than direct query execution, frameworks like InPars~\citep{bonifacio2022inpars} restrict the use of generative text to synthesizing offline training data for neural models. Collectively, these methodologies indicate that synthetic, LLM-generated text aligns far better with latent semantic spaces than with strict term-matching mechanics. Our work provides empirical confirmation of this behavior at production scale within a commercial catalog framework.

\paragraph{LLM Query Expansion Failures.} Generative query expansion is susceptible to specific systemic vulnerabilities, particularly when handling unfamiliar topics or ambiguous source inputs that trigger narrow, biased refinements~\citep{abe2025qe_fails}. Furthermore, rewriting queries without factoring in the specific index capabilities of the underlying retrieval engine can produce fundamentally invalid retrieval plans. To mitigate this issue, probe-based architectures have been proposed to inspect the target index state before executing a plan~\citep{chen2026probe}. Our observation that LLM-generated retargeting queries function as degraded paraphrases of exact keywords serves as a clear instance of this backend-agnostic failure mode when applied to highly structured product catalog fields.

\paragraph{Cost-Aware LLM Routing.} Managing the high computational overhead of generative models is typically addressed via dynamic routing cascades like FrugalGPT~\citep{chen2023frugalgpt}, which uses learned scorers to direct inputs to cheaper or more complex model tiers to achieve up to a 98\% reduction in cost. Similarly, frameworks like RouteLLM~\citep{ong2024routellm} utilize human preference data to train efficient multi-tier routers, while systems like AutoMix~\citep{madaan2023automix} apply self-verification loops to decide when an output requires escalation to a larger foundational model. Operational deployments like Hybrid LLM~\citep{ding2024hybridllm} establish thresholds to split inference workloads between edge devices and cloud infrastructure, and architectures like COEF-VQ~\citep{dong2024coefvq} leverage cheap entropy-based filtering to gate heavy multimodal pipelines. Alternatively, Self-RAG~\citep{asai2024selfrag} introduces internal critique tokens to dictate exactly when external knowledge retrieval is warranted. While most of these frameworks route between language models of varying sizes based on properties of the input, SMART applies the routing idea at a different decision point: a learned gate audits the \emph{output} of a zero-cost deterministic keyword pass and decides whether invoking a generative engine is warranted at all.

\paragraph{Product Retrieval for Ads.}
DPA systems typically adapt the collaborative filtering and embedding-based candidate-generation paradigms developed for large-scale recommenders~\citep{covington2016youtube}.
Hybrid sparse--dense retrieval is likewise established practice, from unsupervised rank fusion~\citep{cormack2009rrf} to systematic analyses of sparse, dense, and hybrid representations~\citep{luan2021sparse}; we treat the hybrid itself as prior art and focus on when each path helps and whether the expensive path should run at all.
Abutbul et al.~\citep{abutbul2023dpa} describe audience prospecting for DPA at Yahoo using retargeting and search-based approaches without LLM query generation.
Hu et al.~\citep{hu2026adnanny} note that deploying LLMs directly in online ad systems is ``impractical due to millisecond latency constraints,'' using LLMs only for offline tasks.
Pinterest~\citep{wang2024pinterest} integrates LLMs for search relevance scoring but not for query generation.
To our knowledge, no prior work provides a controlled study of when LLM-generated queries help or hurt in production ad retrieval.

%% file: sections/preliminary_study.tex
\section{Preliminary Study: Retargeting vs.\ Prospecting}
\label{sec:preliminary}

Before describing the architecture of SMART, we present a diagnostic production study run on live traffic logs. This empirical analysis exposes the core behavior and limitations of traditional lexical retrieval and generative retrieval paths, defining the requirements for an optimized production setup.

\subsection{Diagnostic Setup and Metrics}
\label{sec:diag_setup}

We sample an offline evaluation group of over 20K unique users from production DPA logs across multiple countries. For each user, their full interaction behavior (browsing, cart, and purchase events) recorded up to day $T-1$ serves as the sole input to query generation. This mirrors the deployed system, where no current-session context is used: query generation consumes the accumulated history alone, while the routing gate conditions on that history, static profile attributes, and the keyword-pass output (Section~\ref{sec:system}). The candidate configurations retrieve the top-200 products from an active catalog of tens of millions of listings, which are then evaluated against the user's actual purchases documented on day $T$ and against an offline LLM judge that emits a Relevance Score. All conditions share identical underlying indices, reranker models, and pipeline layouts to isolate query generation performance. Full implementation parameters are detailed in Section~\ref{sec:setup}. Performance is tracked through two distinct, structurally opposing lenses:
\begin{itemize}[leftmargin=*]
    \item \textbf{CatRecall@$K$} (Retargeting Metric): The fraction of Google Product Categories (GPC) purchased on day $T$ that are successfully surfaced within the top-$K$ retrieved items. Category-level evaluation is matching at the natural granularity for DPA systems because exact product-ID recall is vanishingly sparse at scale.
    \item \textbf{Relevance Score} (Prospecting Metric): An LLM-as-a-judge~\citep{zheng2024judging} evaluation using Gemini~3 Flash~\citep{team2024gemini} (with thinking mode enabled) scores the top-10 retrieved items against the user's full shopping history on a 0--3 scale ($3 = $ highly relevant, $2 = $ matches intent, $1 = \text{adjacent}$, $0 = \text{irrelevant}$). This metric explicitly captures semantic alignment to broader intent, rewarding discovery within categories the user has not yet purchased.
\end{itemize}

We evaluate two distinct query strategies: \textit{rule-generated keywords} and \textit{LLM-generated queries} across two retrieval backends: a Tantivy-based BM25 lexical index and a dense FAISS-based Approximate Nearest Neighbor (ANN) index.

\subsection{Token-Frequency Limits on Lexical Search}
\label{sec:keyword_vs_llm}

We first cross-examine the behavior of both query strategies on a shared BM25 index to isolate the characteristics of pure lexical matching.

\begin{table}[t]
\centering
\caption{Lexical retrieval comparison on BM25. Rule-generated keyword queries achieve higher CatRecall@$K$ at every evaluation depth, reflecting their structural alignment with token-frequency matching.}
\label{tab:bm25_headtohead}
\small
\begin{tabular}{@{}lccc@{}}
\toprule
$K$ & Keyword (baseline) & LLM retarget+prospect & $\Delta$ rel. \\
\midrule
5 & 41.0\% & 34.5\% & $-$15.9\% \\
20 & 44.4\% & 37.1\% & $-$16.4\% \\
200 & 56.0\% & 42.5\% & $-$24.1\% \\
\bottomrule
\end{tabular}
\end{table}

Table~\ref{tab:bm25_headtohead} reveals an explicit performance gap: the deterministic Keyword baseline outperforms the generative LLM strategy across every depth, securing a $+15.9\%$ relative advantage at $K=5$ and a $+24.1\%$ lead at $K=200$. This behavior is a direct consequence of index mechanics: BM25 functions as a token-frequency matcher that weights fields by inverse document frequency. Rule-generated queries emit exact brand and category strings drawn directly from user text logs, matching indexed product metadata exactly. When an LLM paraphrases this precise intent (e.g., expanding a brand title into natural description strings), it introduces vocabulary permutations that do not exist verbatim within catalog fields, effectively diluting term frequency and degrading lexical precision. Thus, rule-generated keywords are superior for exact retargeting objectives.

\subsection{Cross-Metric Index Inversion}
\label{sec:judge_inv}

The narrative completely reverses when moving candidates into a dense vector space. Table~\ref{tab:cross_metric} maps both query generation styles across both the BM25 and dense ANN indices evaluated across both key metrics.

\begin{table}[t]
\centering
\caption{Cross-metric backend comparison. CatRecall is evaluated at $K=200$; Relevance Score is checked at $K=10$ on a 0--3 scale. Rule-generated keyword + BM25 leads on CatRecall (retargeting), while LLM + ANN dominates on Relevance Score (prospecting).}
\label{tab:cross_metric}
\small
\begin{tabular}{@{}lcc@{}}
\toprule
System & CatRecall@200 ($\Delta$ rel.) & Rel. Score ($\Delta$ rel.) \\
\midrule
Keyword $+$ BM25 (base) & \textbf{56.0\%} & 1.95 \\
Keyword $+$ ANN & 53.9\% \,($-$3.8\%) & 2.25 \,($+$15.3\%) \\
LLM $+$ BM25 & 42.5\% \,($-$24.1\%) & 2.32 \,($+$18.9\%) \\
\textbf{LLM $+$ ANN} & 50.7\% \,($-$9.5\%) & \textbf{2.39 \,($+$22.6\%)} \\
\bottomrule
\end{tabular}
\end{table}

We observe a clear cross-metric inversion: LLM queries improve Relevance Score across any backend ($+18.9\%$ on BM25, $+22.6\%$ on ANN), yet the keyword rules cleanly dominate on CatRecall@200. Crucially, each query style is best served by its matched backend: switching from BM25 to ANN boosts CatRecall@200 by $+19.3\%$ relative for the LLM path, while the same switch costs keyword queries $3.8\%$. Because dense indices reward conceptual vector proximity over strict token matching, they enable the LLM queries to exploit pre-trained world knowledge, successfully surfacing highly relevant, out-of-history discovery items (prospecting). These findings confirm that keywords + BM25 and LLM + ANN serve completely different retrieval objectives.

\subsection{Quantifying Gating Opportunities}
\label{sec:who_benefits}

While running both paths simultaneously for all incoming traffic would technically maximize candidate pools, two production constraints make a universal configuration impossible. First, global LLM inference introduces massive computational costs and violates strict millisecond ad serving response budgets. Second, our user-level impact analysis (Table~\ref{tab:exploration_impact}) indicates that the benefits of semantic prospecting are heavily concentrated.

\begin{table}[t]
\centering
\caption{Per-user impact of adding LLM queries to retrieval. Only 14.4\% of users see clear metric improvement, while 73\% achieve optimal coverage via exact keywords alone.}
\label{tab:exploration_impact}
\small
\begin{tabular}{@{}lc@{}}
\toprule
Outcome & \% of users \\
\midrule
Both CatRecall@200 and Relevance Score & 2.6\% \\
CatRecall@200 only & 2.2\% \\
Relevance Score only & 9.6\% \\
\textbf{Either metric (helped)} & \textbf{14.4\%} \\
No change & 73.0\% \\
Declined on either metric & 12.6\% \\
\bottomrule
\end{tabular}
\end{table}

A striking $73.0\%$ of users experience absolutely no metric change when generative expansion is appended, as their intent is already completely satisfied by the cheap keyword path. Processing universal traffic through an LLM would yield massive infrastructure waste and introduce candidate noise for the majority of requests. Nor can the downstream reranker fully absorb this noise: it demotes weakly matching items within the merged pool, but candidates displaced from the finite retrieval pool are unrecoverable---indeed, indiscriminate LLM augmentation lowers CatRecall@200 even after reranking (Table~\ref{tab:cost}). This structural asymmetry motivates the proposed architecture: a framework that routes all traffic through the zero-cost keyword path, paired with an adaptive gate that conditionally invokes the expensive LLM vector path only for the targeted $\sim$10\% of users facing a candidate coverage gap.

%% file: sections/architecture.tex
\section{The SMART System Architecture}
\label{sec:system}

The SMART system orchestrates a dual candidate generation loop managed by an online output-auditing routing layer, matching each query strategy to its empirically optimized index backend, as Figure~\ref{fig:architecture} shows.

\subsection{Multi-Stage Retrieval Pipeline}
\label{sec:pipeline}

The core pipeline processes incoming requests across four sequential operational stages:

\paragraph{1. Query Generation.} For a user history $H_u = \{(e_i, t_i, p_i)\}$, SMART combines a deterministic keyword query generator with a selective generative path.
Algorithm~\ref{alg:keyword} describes the rule-generated keyword component, which is explicitly constructed to be backend-aware, emitting tokens guaranteed to appear within the BM25 catalog fields. It tracks brand engagement by computing a score $s(b) = \sum w(e_i)$, where weights prioritize down-funnel conversions ($w(\text{Purchase}) = 3$, $w(\text{Add-to-Cart}) = 2$, $w(\text{View}) = 1$). The top $n/2$ brands by descending engagement are paired with their most frequent historical leaf categories to form exact search tokens, and remaining slots are filled with high-frequency category strings to broaden baseline coverage.

\begin{algorithm}[t]
\caption{Backend-Aware Keyword Query Generation}
\label{alg:keyword}
\begin{algorithmic}[1]
\REQUIRE User history $H_u = \{(e_i, t_i, p_i)\}$, budget $n$ queries
\ENSURE Query set $Q_{\text{kw}}$
\FOR{each brand $b$ in $H_u$}
    \STATE $s(b) \gets \sum_{e_i : \text{brand}(p_i) = b} w(e_i)$ \COMMENT{Purchase=3, Add-to-cart=2, View=1}
\ENDFOR
\STATE $B \gets \text{top-}(n/2)$ brands by $s(b)$ descending
\FOR{each $b \in B$}
    \STATE $c_b \gets$ most frequent leaf category for $b$ in $H_u$
    \STATE $Q_{\text{kw}} \gets Q_{\text{kw}} \cup \{b \oplus c_b\}$ \COMMENT{e.g., ``Brand-X Running Shoes''}
\ENDFOR
\STATE $C \gets$ top-$(n - |Q_{\text{kw}}|)$ categories by frequency, excluding covered
\FOR{each $c \in C$}
    \STATE $Q_{\text{kw}} \gets Q_{\text{kw}} \cup \{c\}$ \COMMENT{e.g., ``Baby Clothing''}
\ENDFOR
\RETURN $Q_{\text{kw}}$
\end{algorithmic}
\end{algorithm}

\paragraph{2. Retrieval.} Keyword queries run against a Tantivy~\citep{tantivy2023} BM25~\citep{robertson2009bm25} index built from the catalog.
Each product document is indexed with title, brand, Google Product Category, product type, and a 5-level hierarchical semantic ID following the TIGER framework~\citep{rajput2023tiger}.
The semantic IDs are generated by fine-tuning a SigLIP2~\citep{tschannen2025siglip2} multimodal vision-language encoder on the Snap product catalog to produce product embeddings, then applying Residual Quantization $k$-means (RQ-KMeans) to discretize the continuous embedding space into a 5-level hierarchy---L1 captures broad product domains (e.g., ``Electronics''), while L5 identifies near-duplicate products.
When selectively invoked by the quality gate, semantic prospecting queries are processed via a single LLM call and executed against a dense ANN index managed via FAISS~\citep{douze2024faiss}.

\paragraph{3. Reranking.} Candidates retrieved from both paths are merged and scored using a LambdaMART~\citep{burges2010lambdamart} learning-to-rank model trained with LightGBM~\citep{ke2017lightgbm}, followed by a Maximal Marginal Relevance (MMR)~\citep{carbonell1998mmr} layer to enforce brand, category, and semantic diversity.

\paragraph{4. Ad Serving.} Top-ranked products are mapped to active advertiser product sets to ensure absolute alignment with live ad campaigns before serving.

\subsection{Adaptive Quality Gate Mechanism}
\label{sec:gate_mech}

To bypass expensive generative processing for users already well-served by rule generated keywords, SMART places an adaptive classifier directly after the zero-cost keyword retrieval pass, deciding whether to invoke the generative prospecting path, as Algorithm~\ref{alg:routing} describes.

\begin{algorithm}[t]
\caption{Adaptive Query Routing}
\label{alg:routing}
\begin{algorithmic}[1]
\REQUIRE User history $H_u$, gate model $g$, threshold $\theta$
\ENSURE Retrieved product set $\mathcal{R}(u)$
\STATE $Q_{\text{kw}} \gets \textsc{KeywordQueries}(H_u)$ \COMMENT{Algorithm~\ref{alg:keyword}, zero cost}
\STATE $\mathcal{R}_{\text{kw}} \gets \textsc{BM25Search}(Q_{\text{kw}})$
\STATE $\mathbf{x} \gets f(H_u, \mathcal{R}_{\text{kw}})$ \COMMENT{Coverage gap features}
\IF{$g(\mathbf{x}) \geq \theta$}
    \STATE $Q_{\text{llm}} \gets \textsc{LLM}(H_u, Q_{\text{kw}})$ \COMMENT{coverage gap: prospecting queries}
    \STATE $\mathcal{R}_{\text{llm}} \gets \textsc{ANNSearch}(Q_{\text{llm}})$
    \STATE $\mathcal{R}(u) \gets \textsc{Rerank}(\mathcal{R}_{\text{kw}} \cup \mathcal{R}_{\text{llm}})$
\ELSE
    \STATE $\mathcal{R}(u) \gets \textsc{Rerank}(\mathcal{R}_{\text{kw}})$
\ENDIF
\RETURN $\mathcal{R}(u)$
\end{algorithmic}
\end{algorithm}

\paragraph{Problem Formulation.} Let $\mathcal{R}_{\text{kw}}(u)$ denote candidates retrieved via the zero-cost keyword pass, and $\mathcal{R}_{\text{smart}}(u)$ denote the combined multi-backend candidate pool. The incremental retrieval benefit $\Delta(u)$ is defined as:
\begin{equation}
    \Delta(u) = m(\mathcal{R}_{\text{smart}}(u)) - m(\mathcal{R}_{\text{kw}}(u))
\end{equation}
where $m(\cdot)$ represents retrieval quality via category recall. The routing objective maximizes global retrieval quality over all users subject to a strict maximum allowable LLM budget fraction $B$:
\begin{equation}
    \max_{\theta} \sum_{u \in \mathcal{U}} m\!\left(\mathcal{R}_\theta(u)\right) \quad \text{s.t.} \quad \frac{|\{u : g(u) \geq \theta\}|}{|\mathcal{U}|} \leq B
\end{equation}
where $\mathcal{R}_\theta(u) = \mathcal{R}_{\text{smart}}(u)$ if the gate score $g(u) \geq \theta$, and $\mathcal{R}_{\text{kw}}(u)$ otherwise.

\paragraph{Gate Architecture and Feature Engineering.} The gate functions as a lightweight binary GBDT classifier~\citep{ke2017lightgbm} evaluating 54 real-time features across 11 structural groups (summarized in Table~\ref{tab:gate_features}). Crucially, rather than relying solely on static user histories, the gate conditions its routing choice directly on the \emph{output insufficiency} of the cheap keyword pass by measuring the structural coverage gap:
\begin{equation}
    \text{gap}(u) = 1 - \frac{|\text{cat}(\mathcal{R}_{\text{kw}}(u)) \cap \text{cat}(H_u)|}{|\text{cat}(H_u)|}
\end{equation}
where $\text{cat}(\cdot)$ maps retrieved items to their unique category taxonomies. A high gap indicates that a user's diverse historical affinities have been displaced or unfulfilled within the top lexical matches, marking a requirement for generative prospecting. This output-auditing design distinguishes the gate from input-based cost-aware routers~\citep{chen2023frugalgpt, ong2024routellm}, which decide from the input alone before any retrieval is attempted: here the router observes what the cheap pass actually retrieved before spending on the expensive path. Training labels are mapped as positive via an OR-of-ORs construction if SMART beats keywords on CatRecall@200 across either of two independent cycles, capturing stable lift potential while minimizing LLM stochasticity.

\begin{table}[t]
\centering
\caption{Quality gate feature groups. Bold groups are derived directly from the real-time keyword search results and serve as the most discriminative signals. Semantic L1/L2 cluster IDs are produced by hierarchically clustering SigLIP2 vision--language embeddings of catalog products.}
\label{tab:gate_features}
\small
\begin{tabular}{@{}llc@{}}
\toprule
Group & Key Features & Count \\
\midrule
User history & Brand/category entropy, diversity ratio & 15 \\
Temporal & History span, events/day, recency & 5 \\
Purchase value & Avg/max purchase value & 3 \\
Match quality & Avg/min BM25 match scores & 2 \\
Category depth & GPC depth, L1 category count & 3 \\
Semantic history & Semantic L1/L2 cluster counts & 3 \\
Brand--category & Categories per brand, brands per cat & 2 \\
Query features & Query count, avg tokens, brand queries & 3 \\
\textbf{Search results} & \textbf{Products, categories, coverage} & \textbf{7} \\
\textbf{Coverage gaps} & \textbf{Category/semantic gap, diversity} & \textbf{9} \\
Profile & Age group, gender & 2 \\
\bottomrule
\end{tabular}
\end{table}

\subsection{Core Implementation Specifications}
\label{sec:setup}

The candidate generation infrastructure runs against a Tantivy BM25 index built from the active catalog. The generative path utilizes Gemini 2.5 Flash Lite via a single structured API call to output exactly three prospecting queries; retargeting is handled entirely by the rule-generated keyword path. The lightweight Flash Lite model is chosen for the cost-sensitive production generation path, whereas the heavier Gemini~3 Flash (thinking mode) serves only as the offline evaluation judge (Section~\ref{sec:diag_setup}). Conceptual representations are encoded using the 768-dimension \texttt{gemini-embedding-001} model over a concatenated text string of product title, brand, description, and category attributes. Dense retrieval is executed over vector fields mapped via FAISS GPU IVFFlat.

The LambdaMART reranker orders candidates utilizing 55 unique user, product, and cross-backend matching features. It targets a graded supervision label ranging from 0 to 15 (distinct from the LLM-judge Relevance Score), calculated by combining: (i) semantic prefix-match length across 5 tiers of the same hierarchical SigLIP2 embedding used to derive the gate's Semantic L1/L2 features (Table~\ref{tab:gate_features}), and (ii) down-funnel engagement interaction weights ($\text{purchase} = 3$, $\text{add-to-cart} = 2$, $\text{view} = 1$). This dense supervision methodology yields $50\times$ more positive training signals than strict item-ID matching while natively isolating high-intent items.

%% file: sections/evaluation.tex
\section{Experimental Evaluation and Ablations}
\label{sec:evaluation}

We evaluate SMART end-to-end, validating the routing gate accuracy, system stability, qualitative behaviors, and production live-traffic performance.

\subsection{Routing Gate Ablation Studies}
\label{sec:routing}

\paragraph{Feature Importance Calculations.} We train a gradient-boosted gate model to analyze feature contributions. Table~\ref{tab:gate_importance} tracks relative information gain across top-performing features.

\begin{table}[t]
\centering
\caption{Top gate features by gain. The classifier accurately learns to identify users whose keyword results yield poor coverage of historical interests.}
\label{tab:gate_importance}
\small
\begin{tabular}{@{}clcl@{}}
\toprule
Rank & Feature & Rel.\ gain & Source \\
\midrule
1 & Product set density & 1.00 & Search results \\
2 & Avg.\ query token count & 0.53 & Query \\
3 & Result leaf categories & 0.30 & Search results \\
4 & Semantic L1 coverage & 0.27 & Coverage gap \\
5 & Brands per category & 0.24 & Brand--category \\
6 & Result diversity & 0.20 & Coverage gaps \\
7 & Avg.\ purchase value & 0.16 & Purchase value \\
\bottomrule
\end{tabular}
\end{table}

The empirical distribution confirms that search-result features (Product set density, gain = 1.00) and explicit coverage gap metrics (Semantic L1 coverage, gain = 0.27) represent the primary predictive signals, justifying our output-auditing routing design over static entry-history routing.

\begin{figure}[t]
    \centering
    \includegraphics[width=0.9\linewidth]{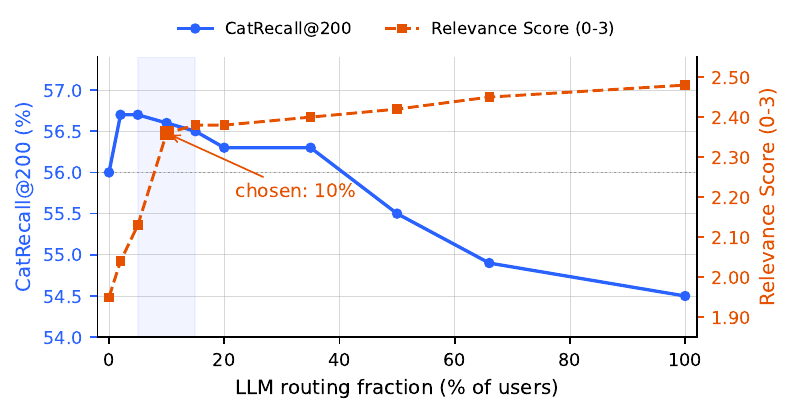}
    \caption{Quality--cost trade-off across LLM routing fraction. CatRecall@200 (blue) peaks within the 5--15\% band, while Relevance Score (orange dashed) demonstrates steep, immediate gains within that identical operating range.}
    \label{fig:routing_sweep}
\end{figure}

\paragraph{Operational Quality-Cost Threshold Sweeps.} We sweep the gate threshold $\theta$ from 0 (always-SMART) to $\infty$ (keyword-only), mapping performance trajectories across varying LLM routing fractions in Figure~\ref{fig:routing_sweep}. 

CatRecall@200 demonstrates an inverse-U curve, peaking at $56.6\%$ within the $10\%$ routing band before decaying as further uncalibrated LLM candidates introduce pool dilution and noise. Concurrently, Relevance Score exhibits its steepest trajectory gains within the same 5--15\% band, encountering clear diminishing returns thereafter. Setting the gate operating threshold to route exactly $10\%$ of incoming traffic captures the peak CatRecall@200 and the bulk of semantic prospecting gains, lifting Relevance Score from $1.95$ (keyword-only) to $2.36$---approaching the $2.39$ of the standalone LLM\,$+$\,ANN configuration (Table~\ref{tab:cross_metric}); the recall--cost operating points are summarized in Table~\ref{tab:cost}.

\begin{table}[t]
\centering
\caption{Operating points at production scale. Gated 10\% achieves optimal category recall at a strict 90\% LLM infrastructure cost reduction versus always-on routing. Deltas are reported relative to the keyword baseline.}
\label{tab:cost}
\small
\begin{tabular}{@{}lcccc@{}}
\toprule
Strategy & LLM\% & CatRecall@200 & $\Delta$ rel. & Rel.\ cost \\
\midrule
Keyword-only (baseline) & 0\% & 56.0\% & --- & 0$\times$ \\
\textbf{Gated 10\%} & \textbf{10\%} & \textbf{56.6\%} & \textbf{$+$1.1\%} & \textbf{0.1$\times$} \\
Gated 15\% & 15\% & 56.5\% & $+$0.9\% & 0.15$\times$ \\
Always SMART & 100\% & 54.5\% & $-$2.7\% & 1$\times$ \\
LLM retarget+prospect & 100\% & 42.5\% & $-$24.1\% & 1$\times$ \\
\bottomrule
\end{tabular}
\end{table}

\subsection{Qualitative Case Studies}
\label{sec:qualitative}

Table~\ref{tab:qualitative} details end-to-end walkthroughs across three distinct consumer profiles, demonstrating how the dual-backend layout resolves systemic failure modes of single-backend configurations.

\begin{table*}[t]
\centering
\caption{Three end-to-end pipeline walkthroughs (advertiser brands anonymized by category role). Each row illustrates a distinct failure mode of single-backend search; SMART resolves all three structurally.}
\label{tab:qualitative}
\footnotesize
\begin{tabular}{@{}p{0.16\linewidth}p{0.38\linewidth}p{0.38\linewidth}@{}}
\toprule
User profile & KW queries $\to$ BM25 top-3 & LLM queries $\to$ ANN top-3 \\
\midrule
\textbf{User A}: mostly fashion $+$ jewelry; tail browsing (baseball gear, phone jammers, 2--5 events each) &
\emph{Queries}: brand $+$ category for her three main brands, plus tail-noise queries (``baseball helmets'', ``phone jammers'').\newline
\emph{Top-3}: three sporting-goods batting helmets/gloves.\newline
\emph{Failure}: tail query monopolizes BM25; her main apparel and jewelry brands fall below rank 10. &
\emph{Queries}: ``women's dress pants'' and ``silver rings'' targeted at her main apparel and jewelry brands, plus persona-grounded paraphrases (``minimalist makeup brush set'').\newline
\emph{Top-3}: three of her main-brand dress pants in different styles.\newline
\emph{Outcome}: main brand recovered; persona-inferred beauty added. \\
\midrule
\textbf{User B}: dominant signal marketplace books; secondary signal two Indian ethnic-wear brands &
\emph{Queries}: each brand name verbatim, plus ``Print Books''.\newline
\emph{Top-3}: three marketplace book listings (children's, philosophy, language-learning).\newline
\emph{Outcome}: her dominant books signal is captured at the top of the ranking. &
\emph{Queries}: brand $+$ ethnic-wear vocabulary (``kurta'', ``palazzo pants''), plus discovery paraphrases (``ethnic wear kurtis'').\newline
\emph{Top-3}: wide-leg printed palazzo pants from three boho-fashion brands.\newline
\emph{Outcome}: brand-diverse ethnic-wear surface; books discarded. \\
\midrule
\textbf{User C}: main brands one fast-fashion $+$ one activewear; ``Beauty Maven'' persona (no K-beauty in clicks); tail baby-clothing events &
\emph{Queries}: brand $+$ category for her two main brands, plus ``Baby \& Toddler Clothing''.\newline
\emph{Top-3}: three toddler sleep bags from a single baby-clothing advertiser.\newline
\emph{Failure}: tail baby query dominates BM25; her main brands fall below rank 10. &
\emph{Queries}: activewear paraphrases for her main brand, plus persona-grounded discovery (``serum for dark spots'', ``SPF 50 sunscreen'', ``hair mask for dry hair'').\newline
\emph{Top-3}: three of her main-brand activewear tops.\newline
\emph{Outcome}: main brand recovered; three new K-beauty brands surfaced via persona expansion. \\
\bottomrule
\end{tabular}
\end{table*}

The pattern is highly consistent: LLM queries append clear semantic precision (User C's K-beauty surfacing), yet their focus on matching a coherent persona causes them to drop loose, unaligned affinities like User B's marketplace book purchases. Conversely, keywords + BM25 emits search queries for all historical interactions at zero cost, preserving coverage for tail or high-volume affinities even when displaced from the top slots. SMART's structural pairing---utilizing the keyword rules as the history-faithful candidate backbone (170 of 200 slots) and the LLM dense path to populate the top ranking layer (30 slots)---mitigates this trade-off in the illustrated cases.

\subsection{System Generalization and Stability}
\label{sec:gen_stab}

\paragraph{Run-to-Run Stability Verification.} To ensure the gate captures stable intent rather than artifacts of LLM stochasticity, we execute parallel generation cycles on identical user logs. Table~\ref{tab:stability} tracks cross-run alignment.

\begin{table}[t]
\centering
\caption{Run-to-run stability of LLM impact. 80.1\% of users showing positive metric gains in the initial run maintain positive trajectories in the validation run, confirming the signal is robust.}
\label{tab:stability}
\small
\begin{tabular}{@{}lccc@{}}
\toprule
Run 1 outcome & Still same & Flipped & Neutral \\
\midrule
Improved & 80.1\% & 5.1\% & 14.8\% \\
Declined on CatRecall@200 & 75.6\% & 2.2\% & 22.2\% \\
\bottomrule
\end{tabular}
\end{table}

The consistency is strong: $80.1\%$ of users improved in Run 1 are stably improved in Run 2, confirming that user-level benefit from generative expansion is an invariant, learnable property of the request state.

\paragraph{Temporal Evaluation Cross-Validation.} We evaluate gate generalization capacity under a strict temporal protocol: models are trained on day $T$ and tested on completely non-overlapping, held-out day $T+1$ user sets across distinct catalog states.

\begin{table}[t]
\centering
\caption{Temporal generalization: gate models trained on day $T$, evaluated on held-out day $T+1$ user sets.}
\label{tab:temporal}
\small
\begin{tabular}{@{}lcc@{}}
\toprule
Model & AUC & CatRecall@5 $\Delta$ \\
\midrule
GBDT & 0.753 & +0.99pp \\
DNN & 0.748 & +1.13pp \\
\midrule
Oracle (per-user) & --- & +6.9pp \\
\bottomrule
\end{tabular}
\end{table}

Table~\ref{tab:temporal} shows that the gate generalizes across time. GBDT achieves an AUC of $0.753$ on unseen-day users, driving a $+0.99\text{pp}$ CatRecall@5 lift over baseline. A deep neural network (DNN) variant yields comparable metrics ($\text{AUC} = 0.748$, $+1.13\text{pp}$ lift), demonstrating that the coverage-gap predictive signal is robust and model-agnostic.

\begin{figure}[t]
    \centering
    \includegraphics[width=0.95\linewidth]{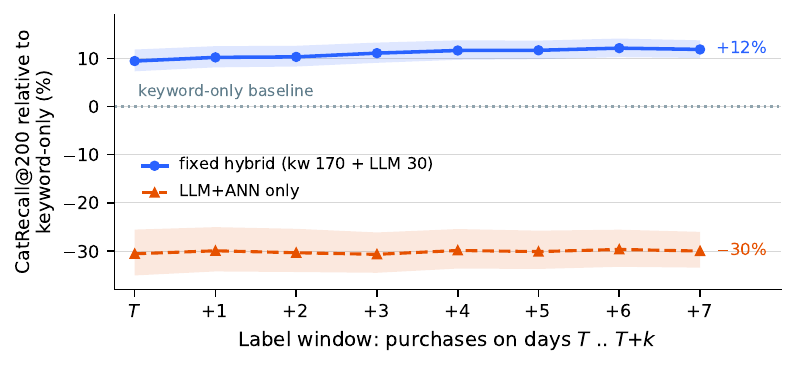}
    \caption{Sensitivity of CatRecall@200 to the purchase-label horizon. Per user, history, catalog snapshot, and retrieved top-200 candidates are fixed at day $T$; only the cumulative purchase-category label window expands from $T$ to $T$..$T{+}k$. Curves show lift over keyword-only ($=0$) with pointwise 95\% user-bootstrap intervals, on a converter-conditioned cohort. The ordering is stable across all horizons.}
    \label{fig:lag_composition}
\end{figure}

\paragraph{Robustness to Delayed Purchase Labels.}A relevant category may convert only after a delay---over half of the categories a user buys in the week after day $T$ are absent from the day-$T$ label set---so we check whether our single-day CatRecall labels bias the comparison. Holding each user's history, catalog snapshot, and retrieved top-200 candidates fixed, we expand only the evaluated labels from day $T$ to day $T$ plus the following seven days, and re-score each configuration---all retrieving 200 candidates---relative to keyword-only (Figure~\ref{fig:lag_composition}). The day-$T$ ordering is unchanged across horizons: the fixed keyword\,$+$\,LLM hybrid (170\,$+$\,30 candidates) holds a stable $+$9--12\% edge over keyword-only (pointwise 95\% bootstrap intervals exclude zero at every horizon), while LLM$+$ANN alone stays well below it (consistent with Section~\ref{sec:judge_inv}). This converter-conditioned cohort uses the production catalog and candidate budget, so its recall is not directly comparable to Tables~\ref{tab:cross_metric} and~\ref{tab:cost}; we read it as a sensitivity check for delayed labels rather than a prospecting measure---longer windows reduce label censoring but admit post-$T$ intent drift, and never-purchased relevant items stay unlabeled, leaving Relevance Score the complementary prospecting metric.

\subsection{Online Production Deployment}
\label{sec:deployment}

We validate the framework through a 2-week online A/B test on live Snap traffic against an optimized production baseline consisting of a two-tower user-to-product (U2I) ANN retrieval model evaluating hundreds of features. Production traffic was split uniformly at random between arms. This test also supplies the comparison against a state-of-the-art retriever that our offline studies intentionally omit: Sections~\ref{sec:preliminary} and~\ref{sec:routing}--\ref{sec:gen_stab} hold indices and rerankers fixed to isolate query-generation effects, whereas here SMART is measured head-to-head against the production embedding-based model under live traffic.

\begin{table}[t]
\centering
\caption{Online A/B production test results (2 weeks). SMART is evaluated directly against Snap's production embedding-based U2I framework.}
\label{tab:online_ab}
\small
\begin{tabular}{@{}lcc@{}}
\toprule
Method & CR & CTR \\
\midrule
Embedding-based U2I (baseline) & 100\% & 100\% \\
SMART (ours) & \textbf{127.6\%} & 99.1\% \\
\bottomrule
\end{tabular}
\end{table}

Table~\ref{tab:online_ab} logs the live results. SMART delivers a substantial $+27.6\%$ relative improvement in core Conversion Rate (CR), defined as attributed click-and-purchase actions, while Click-Through Rate (CTR) remains flat at $-0.9\%$. This performance divergence confirms that SMART is not generating low-intent clicks indiscriminately, but rather surfaces highly relevant discovery items that convert while actively filtering out low-intent tail noise.

%% file: sections/closing.tex
\section{Discussion, Limitations, and Conclusion}
\label{sec:closing}

\subsection{Discussion}
\label{sec:discussion}

\paragraph{Domain Structural Mismatches.} Prior literature on LLM query expansion~\citep{wang2023query2doc, gao2022hyde, jagerman2023query} focuses almost exclusively on long-form, unstructured documents within academic benchmarks like MS-MARCO. Our findings prove that within commercial catalogs characterized by concise, highly structured titles (5--15 tokens), the lexical mismatch introduced by fluid natural-language LLM paraphrasing is significantly more severe. Generative queries do not function as drop-in replacements for traditional search heuristics; instead, they shift candidate retrieval parameters away from strict historical retargeting toward intent-driven semantic discovery. SMART explicitly accounts for this by matching each query style to its winning index engine.

\paragraph{Capability Selection via Output Gating.} Prevailing cost-aware routing networks (e.g., FrugalGPT~\citep{chen2023frugalgpt}, RouteLLM~\citep{ong2024routellm}) concentrate on \emph{model selection}, analyzing raw input text difficulty to split queries between cheaper and larger LLM variants. SMART instead operates on \emph{capability selection}. By executing a zero-cost heuristic pass first and auditing its structural output for active coverage gaps, the gate conditionally decides whether to invoke generative machinery at all. This execution pattern generalizes to any high-throughput retrieval workflow combining fast heuristics with expensive neural augmentations.

\paragraph{Generalizability.}The transferable core of SMART is the process: decomposing retrieval into retargeting- and prospecting-style objectives, a diagnostic that quantifies who benefits from expensive augmentation, and a gate that audits cheap results before invoking it. The concrete instantiation, however, is catalog-bound: rule-generated queries assume structured brand and category fields, and many gate features derive from category taxonomies or from the semantic-ID hierarchy---though the latter is learned from item embeddings and could be rebuilt in any domain with embeddable items. Practitioners in domains without such structure would need to substitute their own coverage measures over whatever cheap first-pass retrieval exists, while the audit-then-invoke principle of the preceding paragraph carries over unchanged.

\subsection{Limitations and Future Horizons}
\label{sec:limitations}

Several research vectors remain open. First, our temporal validation is currently bounded by next-day holdout testing; extending this tracking across multi-week production horizons will verify systemic robustness against long-term catalog churn and seasonal shopping shifts. Relatedly, the delayed-label robustness check of Section~\ref{sec:gen_stab} extends labels by only one week; longer horizons and attribution of online conversions seeded by each retrieval path remain future work. Second, our offline Relevance Score evaluations utilize a single judge family (Gemini 3 Flash). While we anchor its outputs against real-world conversion metrics to mitigate bias, future configurations will introduce multi-model consensus parsing and human validation panels to eliminate potential model sycophancy. Third, the routing threshold is not load-aware: traffic surges raise LLM call volume proportionally, though service degrades gracefully since every request already completes the keyword pass. Finally, transitioning our binary routing classifier into a continuous regression formulation that directly targets the expected metric delta $\Delta(u)$ will allow for granular, real-time threshold calibration to balance infrastructure cost against fluctuating ad request volume.

\subsection{Conclusion}
\label{sec:conclusion}

The development and deployment of SMART demonstrates that rule-generated keywords and generative language queries operate as structural complements rather than substitutes in production recommender systems. Deterministic keyword rules exploit inverted index token matching to serve as an optimal candidate foundation for retargeting on BM25. Concurrently, generative language models interpret multi-event intent sequences to perform semantic prospecting over dense ANN vector spaces. By bifurcating retrieval paths through an output-aware quality gate, SMART captures the broad discovery capabilities of LLMs, improving online ad conversions by $+27.6\%$ at a strict $90\%$ reduction in total generative infrastructure cost.